\documentclass[pre,aps,preprint,showpacs,preprintnumbers,amsmath,amssymb,secnumroman,eqsecnum]{revtex4}

\usepackage{graphicx}% Include figure files
\usepackage{dcolumn}% Align table columns on decimal point
\usepackage{bm}% bold math

\newcommand{\beq}{\begin{equation}}
\newcommand{\eeq}{\end{equation}}
\newcommand{\beqa}{\begin{eqnarray}}
\newcommand{\eeqa}{\end{eqnarray}}
\newcommand{\vc}[1]{\mbox{\boldmath $#1$}}

\newcommand{\vol}[1]{{\bf #1}}

\newcommand{\du}[1]{{\bf\sf #1}}

\begin{document}

%\preprint{APS/123-QED}

\title{Spinning swimming of Volvox by tangential helical wave}% Force line breaks with \\

\author{B. U. Felderhof}

 \email{ufelder@physik.rwth-aachen.de}
\affiliation{Institut f\"ur Theorie der Statistischen Physik\\ RWTH Aachen University\\
Templergraben 55\\52056 Aachen\\ Germany\\
}%

%\author{R. B. Jones}

 %\email{r.b.jones@qmul.ac.uk}
%\affiliation{Queen Mary University of London, The School of
%Physics and Astronomy, Mile End Road, London E1 4NS, UK\\}%

\date{\today}% It is always \today, today,

\begin{abstract}
The swimming of a sphere by means of tangential helical waves running along its surface is studied on the basis of the Stokes equations. Two types of tangential waves are found. The first of these is associated with a pressure disturbance and leads to a higher rate of net rotation than the second one for the same power. It is suggested that the helical waves are relevant for the rotational swimming of Volvox.
\end{abstract}

\maketitle
\section{\label{I}Introduction}

The protist Volvox consists of a large number of cells embedded in a spherical gelatinous wall. The cells are distributed approximately uniformly on the spherical surface. The interior of the sphere is called the extracellular matrix. The cells have flagella whose collective waving in the external fluid causes locomotion. Antonie van Leeuwenhoek was the first to observe the rotational swimming of Volvox. The name was coined by Linnaeus on the basis of the steady rotation. Gravity provides a direction of reference because of the anisotropy of the inner body. Volvox seems to have a preferential direction of rotation about the axis of gravity \cite{1}.

The effect of waving flagella on the surrounding flow may be modeled as a no-slip boundary condition applied on a periodically deforming surface.
The task of theory is to calculate the translational and rotational swimming velocity and the mean rate of dissipation resulting from assumed periodic surface deformations. In the following we perform such a calculation within the framework of the bilinear theory of swimming \cite{2}-\cite{8}.

We find two dissimilar modes of rotational swimming which may be characterized as tangential helical waves. The two modes lead to opposite directions of steady rotation. One of the modes is associated with a wavelike pressure disturbance. For the same power this mode leads to a larger rate of rotation than the other one. A linear superposition of the two modes can lead to screw-type swimming.

We are dealing with Volvox swimming freely in infinite fluid. In experiments the flow disturbance of Volvox held fixed by a micropipette has been observed \cite{9}-\cite{11}. Fixation significantly alters the flow pattern, since the moments of the surface velocity corresponding to net translation and rotation must be made to vanish. The flow disturbance at some distance from the spherical surface observed by velocimetry in the fixed configuration suggests an axial stroke, as first discussed by Lighthill \cite{2} and Blake \cite{3}. The superposition of axial modes would lead to translational swimming if the Volvox were free. Rotational modes, as discussed above, are possibly present with the same beat frequency but a much shorter wavelength, corresponding to the discrete nature of the cells and the related surface structure.

In Sec. II we discuss the basic equations of the bilinear theory of Stokesian swimming. In Sec. III we study a screw-type spherical swimmer with surface deformations of low multipole order. In Sec. IV we discuss the tangential squirming helical waves and the corresponding swimming velocities and rate of dissipation. In Sec. V we present a tentative analysis of the swimming of Volvox. The article concludes with a discussion.

\section{\label{II}Flow equations}

We consider a sphere of radius $a$ immersed in a viscous
incompressible fluid of shear viscosity $\eta$. At low Reynolds
number and on a slow time scale the flow velocity
$\vc{v}(\vc{r},t)$ and the pressure $p(\vc{r},t)$ satisfy the
Stokes equations
\begin{equation}
\label{2.1}\eta\nabla^2\vc{v}-\nabla p=0,\qquad\nabla\cdot\vc{v}=0.
\end{equation}
The fluid is set in motion by distortions of the
spherical surface which are periodic in time and lead to swimming
motion of the sphere as well as to a time-dependent flow field. The surface displacement
$\vc{\xi}(\vc{s},t)$ is defined as the vector distance
\begin{equation}
\label{2.2}\vc{\xi}=\vc{s}'-\vc{s}
\end{equation}
of a point $\vc{s}'$ on the displaced surface $S(t)$ from the
point $\vc{s}$ on the sphere with surface $S_0$. The fluid
velocity $\vc{v}(\vc{r},t)$ is required to satisfy
\begin{equation}
\label{2.3}\vc{v}(\vc{s}+\vc{\xi}(\vc{s},t))=\frac{\partial\vc{\xi}(\vc{s},t)}{\partial t},
\end{equation}
corresponding to a no-slip boundary condition. The instantaneous translational swimming velocity $\vc{U}(t)$,
the rotational swimming velocity $\vc{\Omega}(t)$, and the flow pattern $(\vc{v},p)$ follow from the condition that no net
force or torque is exerted on the fluid. We evaluate these quantities by a perturbation expansion in powers of the
displacement $\vc{\xi}(\vc{s},t)$.

To second order in $\vc{\xi}$ the flow velocity and the swimming velocity
take the form \cite{6}
\begin{equation}
\label{2.4}\vc{v}(\vc{r},t)=\vc{v}_1(\vc{r},t)+\vc{v}_2(\vc{r},t)+...,\qquad
\vc{U}(t)=\vc{U}_2(t)+....
\end{equation}
Both $\vc{v}_1$ and $\vc{\xi}$ vary harmonically with frequency
$\omega$, and can be expressed as
 \begin{eqnarray}
\label{2.5}\vc{v}_1(\vc{r},t)&=&\vc{v}_{1c}(\vc{r})\cos\omega t+\vc{v}_{1s}(\vc{r})\sin\omega t,\nonumber\\
\vc{\xi}(\vc{s},t)&=&\vc{\xi}_{c}(\vc{s})\cos\omega
t+\vc{\xi}_{s}(\vc{s})\sin\omega t.
\end{eqnarray}
Expanding the no-slip condition Eq. (2.3) to second order we find
for the flow velocity at the surface
\begin{eqnarray}
\label{2.6}\vc{u}_{1S}(\theta,\varphi,t)&=&\vc{v}_1\big|_{r=a}=\frac{\partial\vc{\xi}(\theta,\varphi,t)}{\partial t},\nonumber\\
\vc{u}_{2S}(\theta,\varphi,t)&=&\vc{v}_2\big|_{r=a}=-\vc{\xi}\cdot\nabla\vc{v}_1\big|_{r=a},
\end{eqnarray}
in spherical coordinates $(r,\theta,\varphi)$. In complex notation with $\vc{v}_1=\vc{v}_\omega\exp(-i\omega t)$ the mean second order surface velocity is given by
\begin{equation}
\label{2.7}\overline{\vc{u}}_{2S}(\vc{s})=-\frac{1}{2}\mathrm{Re}(\vc{\xi}^*_\omega\cdot\nabla)\vc{v}_\omega\big|_{r=a},
\end{equation}
where the overhead bar indicates a time-average over a period $T=2\pi/\omega$.

We consider periodic displacements such that the body swims in the $z$ direction. The time-averaged translational swimming velocity is given by \cite{7}
\begin{equation}
\label{2.8}\overline{\vc{U}}_2=\overline{U_2}\;\vc{e}_z,\qquad\overline{U_2}=-\frac{1}{4\pi}\int\overline{\vc{u}}_{2S}\cdot\vc{e}_z\;d\Omega,
\end{equation}
where the integral is over spherical angles $(\theta,\varphi)$.
Similarly the time-averaged rotational swimming velocity is given by  \cite{7}
\begin{equation}
\label{2.9}\overline{\vc{\Omega}}_2=\overline{\Omega_2}\;\vc{e}_z,\qquad\overline{\Omega_2}=-\frac{3}{8\pi a}\int(\vc{e}_r\times\overline{\vc{u}}_{2S})\cdot\vc{e}_z\;d\Omega.
\end{equation}

To second order the rate of dissipation $\mathcal{D}_2(t)$ is
determined entirely by the first order solution. It may be
expressed as a surface integral \cite{6}
\begin{equation}
\label{2.10}\mathcal{D}_2=-\int_{r=a}\vc{v}_{1}\cdot\vc{\sigma}_{1}\cdot\vc{e}_r\;dS,
\end{equation}
where $\vc{\sigma}_1$ is the first order stress tensor, given by
\begin{equation}
\label{2.11}\vc{\sigma}_{1}=\eta(\nabla\vc{v}_1+\widetilde{\nabla\vc{v}_1})-p_1\vc{I}.
\end{equation}
The rate of dissipation is positive and oscillates in time about a
mean value. The mean rate of dissipation equals the power
necessary to generate the motion.

\section{\label{III}Screw-type swimmer}

In our calculations it is convenient to expand the first order flow field and the pressure in terms of a basis set of complex solutions. The general solution of Eq. (2.1) can be expressed as the complex flow velocity and pressure
\begin{eqnarray}
\label{3.1}\vc{v}^c_1(\vc{r},t)&=&-\omega a\sum^\infty_{l=1}\sum^l_{m=-l}\big[\kappa_{lm}\vc{v}_{lm}(\vc{r})+\nu_{lm}\vc{w}_{lm}(\vc{r})+\mu_{lm}\vc{u}_{lm}(\vc{r})\big]e^{-i\omega t},\nonumber\\
p^c_1(\vc{r},t)&=&-\omega a\sum^\infty_{l=1}\sum^l_{m=-l}\kappa_{lm}p_{lm}(\vc{r})e^{-i\omega t},
\end{eqnarray}
with complex coefficients $(\mu_{lm},\kappa_{lm},\nu_{lm})$ and basic solutions \cite{8},\cite{12}
\begin{eqnarray}
\label{3.2}
\vc{v}_{lm}(\vc{r})&=&\bigg(\frac{2l+2}{l(2l+1)}\hat{\vc{A}}_{lm}-\frac{2l-1}{2l+1}\hat{\vc{B}}_{lm}\bigg)\bigg(\frac{a}{r}\bigg)^l,\nonumber\\ p_{lm}(\vc{r})&=&2\eta(2l-1)(-1)^mP^m_l(\cos\theta)e^{im\varphi}\frac{a^l}{r^{l+1}},\nonumber\\
\vc{w}_{lm}(\vc{r})&=&-i\;\hat{\vc{C}}_{lm}\bigg(\frac{a}{r}\bigg)^{l+1},\nonumber\\
\vc{u}_{lm}(\vc{r})&=&-\hat{\vc{B}}_{lm}\bigg(\frac{a}{r}\bigg)^{l+2},
\end{eqnarray}
with vector spherical harmonics $\hat{\vc{A}}_{lm},\hat{\vc{B}}_{lm},\hat{\vc{C}}_{lm}$ in the notation of Ref. 12 (with $2^{l+1}$ in the normalization coefficient replaced by $2l+1$), and with associated Legendre functions $P^m_l$ in the notation of Edmonds \cite{13}. The functions $(\vc{v}_{lm}(\vc{r}), p_{lm}(\vc{r}))$ satisfy the Stokes equations (2.1), and the functions $\vc{u}_{lm}(\vc{r})$ and $\vc{w}_{lm}(\vc{r})$ satisfy these equations with vanishing pressure disturbance. For $m=0$ the solutions are axisymmetric and the functions  $\vc{u}_{lm}(\vc{r}),\vc{v}_{lm}(\vc{r}),p_{lm}(\vc{r})$ are then identical with the functions $\vc{u}_{l}(\vc{r}),\vc{v}_{l}(\vc{r}),p_{l}(\vc{r})$ introduced in Ref. 8. The solutions contain a factor $\exp[i(m\varphi-\omega t)]$, representing a running wave in the azimuthal direction for $m\neq0$. Earlier \cite{12},\cite{14} we have denoted the $\vc{w}_{lm}$- modes as $T$-type, and the $\vc{u}_{lm}$- modes as $P$-type.

We consider first a simple case with linear superpositions of the above solutions with $l=1,2$ and $m=1$. We can exclude the solutions $\vc{v}_{11}(\vc{r})$ and $\vc{w}_{11}(\vc{r})$. In the first the body exerts a force on the fluid, and in the second it exerts a torque. We therefore consider in particular flow situations given by the real part of the expressions
\begin{eqnarray}
\label{3.3}\vc{v}^c_1(\vc{r},t)&=&-\omega a\big[\mu_{11}\vc{u}_{11}(\vc{r})+\kappa_{21}\vc{v}_{21}(\vc{r})+\nu_{21}\vc{w}_{21}(\vc{r})+\mu_{21}\vc{u}_{21}(\vc{r})\big]e^{-i\omega t},\nonumber\\
p^c_1(\vc{r},t)&=&-\omega a\kappa_{21}p_{21}(\vc{r})e^{-i\omega t},
\end{eqnarray}
with four complex coefficients $(\mu_{11},\kappa_{21},\nu_{21},\mu_{21})$. Correspondingly we introduce the complex multipole moment four-vector
\begin{equation}
\label{3.4}\vc{\psi}=(\mu_{11},\kappa_{21},\nu_{21},\mu_{21}).
\end{equation}
Then the mean second order swimming velocity is in the $z$ direction with value $\overline{U_2}$ given by
\begin{equation}
\label{3.5}\overline{U_2}=\frac{1}{2}\;\omega
a(\vc{\psi}|\du{B}|\vc{\psi}),
\end{equation}
with a dimensionless hermitian $4\times 4$ matrix $\du{B}$. The mean second order rotational swimming velocity is in the $z$ direction with value $\overline{\Omega_2}$ given by
\begin{equation}
\label{3.6}\overline{\Omega_2}=\frac{3}{4}\;\omega
(\vc{\psi}|\du{C}|\vc{\psi}),
\end{equation}
with a dimensionless hermitian $4\times 4$ matrix $\du{C}$. The time-averaged rate of dissipation can be expressed as
\begin{equation}
\label{3.7}\overline{\mathcal{D}_2}=8\pi\eta\omega^2a^3(\vc{\psi}|\du{A}|\vc{\psi}),
\end{equation}
with a dimensionless hermitian matrix $\du{A}$. The matrix elements of the three matrices can be evaluated from Eqs. (2.6)-(2.11).

Explicitly we find for the matrix $\du{A}$
\begin{equation}
\label{3.8}\du{A}=\frac{3}{5}\left(\begin{array}{cccc}
10&0&0&0
\\0&27&0&36
\\0&0&12&0
\\0&36&0&60
\end{array}\right),
\end{equation}
for the matrix $\du{B}$
\begin{equation}
\label{3.9}\du{B}=\frac{9}{5}\;i\left(\begin{array}{cccc}
0&-1&0&-5
\\1&0&-3&0
\\0&3&0&5
\\5&0&-5&0
\end{array}\right),
\end{equation}
and for the matrix $\du{C}$
\begin{equation}
\label{3.10}\du{C}=\frac{3}{5}\left(\begin{array}{cccc}
10&0&12&0
\\0&0&0&15
\\12&0&-2&0
\\0&15&0&40
\end{array}\right).
\end{equation}
This is the matrix form of the linear operators for dissipation, translational and rotational swimming in the $VWU$-representation corresponding to the choice of basis set in Eq. (3.1). The limitation to the superposition given in Eq. (3.3) implies that we are considering only a 4-dimensional section of Hilbert space.

We see that the matrix $\du{C}$ is symmetric, unlike $\du{B}$, which is antisymmetric. The nonvanishing diagonal elements imply that a single mode can lead to rotational swimming. In particular the moment vector $(\mu_{11},0,0,0)$, corresponding to a rotating dipole, leads to a nonvanishing mean rotational velocity $\overline{\Omega_2}$. In earlier work \cite{6} we have studied a purely rotational swimmer of this type, with waves corresponding to $m=1$ and $m=-1$ . It was shown there that the mean second order flow velocity and the mean second order pressure disturbance vanish in the laboratory frame (there is a factor $\rho$ missing in Eq.  (11.8) of Ref. 6).

Other rotational swimmers with vanishing $\overline{U_2}$ are provided by the moment vector $(0,0,\nu_{21},0)$ corresponding to a running azimuthal wave, and by the moment vector $(0,0,0,\mu_{21})$ corresponding to a rotating quadrupolar displacement. An example of a moment vector for which both $\overline{U_2}$ and $\overline{\Omega_2}$ are nonvanishing is provided by$(0,\kappa_{21},\nu_{21},0)$ with nonvanishing phase difference between the two coefficients.

Optimization of the mean translational swimming velocity for given mean rate of dissipation leads to the eigenvalue problem
\begin{equation}
\label{3.11}\du{B}|\psi_\lambda)=\lambda\du{A}|\psi_\lambda).
\end{equation}
Both matrices $\du{B}$ and $\du{A}$ are hermitian, so that the eigenvalues $\lambda$ are real. We denote the maximum eigenvalue as $\lambda_{max}$ and the corresponding eigenvector, normalized such that the first component equals unity, as $\vc{\xi}_0$. We find
\begin{equation}
\label{3.12}\lambda_{max}=\sqrt{\frac{1}{96}(65+\sqrt{2305})}\approx 1.085.
\end{equation}
The corresponding eigenvector has components
\begin{equation}
\label{3.13}\vc{\xi}_0\approx (1,-1.024i,-0.361,0.928i).
\end{equation}
The maximum eigenvalue for swimming with axial modes $(\vc{u}_{10},\vc{v}_{20},\vc{u}_{20})$ is $\lambda_{max}=5 /(3\sqrt{2})\approx 1.17851$, showing that with modes up to order $l=2$ the screw-type swimming with modes $(\vc{u}_{11},\vc{v}_{21},\vc{w}_{21},\vc{u}_{21})$ is 9 percent less efficient than axisymmetric swimming.

The instantaneous deformation of the body surface may be chosen to be given by the set of coefficients $\varepsilon\vc{\xi}_0\exp(-i\omega t)$ with amplitude factor $\varepsilon$. In Fig. 1 we show the deformed surface at times $t=0,\;T/8,\;T/4$ for $\varepsilon=0.17$, where $T=2\pi/\omega$ is the period. The values in Eqs. (3.5)-(3.7) corresponding to the eigenvector $\vc{\xi}_0$ are
\begin{equation}
\label{3.14}\overline{U_2}=7.530\varepsilon^2\omega a,\qquad\overline{\Omega_2}=-2.657\varepsilon^2\omega,\qquad\overline{\mathcal{D}_2}=348.862\varepsilon^2\eta\omega^2a^3.
\end{equation}
This shows that the body has a net screw-type solid body motion superposed on the first order screw-type displacements shown in Fig. 1. For $\varepsilon<<1$ the mean rotational velocity $|\overline{\Omega_2}|$ is much smaller than the wave frequency $\omega$.

\section{\label{IV}Tangential squirming swimmer}

 Next we consider squirming swimmers with surface displacement $\vc{\xi}(\vc{s},t)$ tangential to the spherical surface $r=a$. To this purpose it is convenient to make a change of basis. Thus we consider the tangential flow field
\begin{equation}
\label{4.1}\vc{t}_{lm}(\vc{r})=\vc{v}_{lm}(\vc{r})-\vc{u}_{lm}(\vc{r}).
\end{equation}
The associated pressure field is $p_{lm}(\vc{r})$. The fundamental solution $\vc{w}_{lm}(\vc{r})$ is also tangential to the spherical surface $r=a$. We consider displacements $\vc{\xi}(\vc{s},t)$ given by linear combinations of $\vc{t}_{lm}(\vc{s})$ and $\vc{w}_{lm}(\vc{s})$ of the form
\begin{equation}
\label{4.2}\vc{\xi}^c(\vc{s},t)=-i a\big[\tau_{lm}\vc{t}_{lm}(\vc{s})+\nu_{lm}\vc{w}_{lm}(\vc{s})\big]e^{-i\omega t},
\end{equation}
with two complex coefficients $(\tau_{lm},\nu_{lm})$ and corresponding first order flow field
\begin{eqnarray}
\label{4.3}\vc{v}^c_1(\vc{r},t)&=&-\omega a\big[\tau_{lm}\vc{t}_{lm}(\vc{r})+\nu_{lm}\vc{w}_{lm}(\vc{r})\big]e^{-i\omega t},\nonumber\\
p^c_1(\vc{r},t)&=&-\omega a\tau_{lm}p_{lm}(\vc{r})e^{-i\omega t}.
\end{eqnarray}
Correspondingly we introduce the complex moment vector
\begin{equation}
\label{4.4}\vc{\psi}_{lm}=(\tau_{lm},\nu_{lm}).
\end{equation}
We allow integer $l\geq2$ and $m=-l,...,l$.

The basis solutions introduced above are orthogonal in the sense that the corresponding dissipation matrix $\du{A}$ is diagonal in $lm$ subscripts, as given by a factor $\delta_{ll'}\delta_{mm'}$. On the diagonal we find for given $(l,m)$
\begin{eqnarray}
\label{4.5}\du{A}_{lm}=\left(\begin{array}{cc}
f_{lm}&0
\\0&g_{lm}\end{array}\right)
\end{eqnarray}
with elements
\begin{equation}
\label{4.6}f_{lm}=\frac{l+1}{l}\frac{(l+m)!}{(l-m)!},\qquad g_{lm}=\frac{l^2(l+2)}{4(2l+1)}f_{lm}.
\end{equation}
Hence in the chosen subspace of tangential squirming motion the dissipation matrix $\du{A}$ in this representation is diagonal in all subscripts.

For fixed $(l,m)$ we can also consider the  $2\times 2$ matrices $\du{B}_{lm}$ and $\du{C}_{lm}$. In the above representation we find
\begin{eqnarray}
\label{4.7}\du{B}_{lm}=i\left(\begin{array}{cc}
0&u_{lm}
\\-u_{lm}&0\end{array}\right),\qquad\du{C}_{lm}=\left(\begin{array}{cc}
v_{lm}&0
\\0&-w_{lm}\end{array}\right),\qquad
\end{eqnarray}
with elements
\begin{equation}
\label{4.8}u_{lm}=\frac{ml}{2l+1}f_{lm},\qquad v_{lm}=\frac{4(l^2+l-1)}{l^2(l+1)}u_{lm}\qquad w_{lm}=\frac{1}{l+1}u_{lm}.
\end{equation}
The above provides the matrix form of the linear operators for dissipation, translational and rotational swimming in the $TW$-representation corresponding to the choice of basis set $(\vc{t}_{lm},\vc{w}_{lm})$. The limitation to the superposition given in Eq. (4.3) implies that we are considering only a 2-dimensional section of Hilbert space.

The mean translational swimming velocity is given by
\begin{equation}
\label{4.9}\overline{U}_2=\frac{1}{2}\;i(\tau^*_{lm}\nu_{lm}-\tau_{lm}\nu^*_{lm})u_{lm}\omega a,
\end{equation}
the mean rotational swimming velocity is given by
\begin{equation}
\label{4.10}\overline{\Omega}_2=\frac{3}{4}\;\big[|\tau_{lm}|^2v_{lm}-|\nu_{lm}|^2w_{lm}\big]\omega,
\end{equation}
and the mean rate of dissipation is given by
\begin{equation}
\label{4.11}\overline{\mathcal{D}_2}=8\pi\eta\omega^2a^3\big[|\tau_{lm}|^2f_{lm}+|\nu_{lm}|^2g_{lm}\big].
\end{equation}
It is clear from the above expressions that the body spins if just the moment $\tau_{lm}$ or $\nu_{lm}$ is excited. If both $\tau_{lm}$ and $\nu_{lm}$ are nonvanishing, then with proper phase difference there is also a translational swimming velocity. Both swimming velocities vanish for $m=0$ and increase with increasing $|m|$.

We introduce the complex amplitudes
\begin{equation}
\label{4.12}\alpha_{lm+}=\tau_{lm}\sqrt{f_{lm}},\qquad\alpha_{lm-}=\nu_{lm}\sqrt{g_{lm}}.
\end{equation}
With this notation the mean translational swimming velocity becomes
\begin{equation}
\label{4.13}\overline{U}_2=\frac{-im}{\sqrt{(l+2)(2l+1)}}\big(\alpha^*_{lm+}\alpha_{lm-}-\alpha_{lm+}\alpha^*_{lm-}\big)\omega a,
\end{equation}
the mean rotational swimming velocity becomes
\begin{equation}
\label{4.14}\overline{\Omega}_2=\frac{3m}{l(l+1)}\;\bigg(\frac{l^2+l-1}{2l+1}|\alpha_{lm+}|^2-\frac{1}{l+2}|\alpha_{lm-}|^2\bigg)\omega,
\end{equation}
and the mean rate of dissipation becomes
\begin{equation}
\label{4.15}\overline{\mathcal{D}_2}=8\pi\eta\omega^2a^3\big(|\alpha_{lm+}|^2+|\alpha_{lm-}|^2\big).
\end{equation}
We put
\begin{equation}
\label{4.16}\alpha_{lm+}=a_+e^{-i\delta_+},\qquad\alpha_{lm-}=a_-e^{-i\delta_-},
\end{equation}
with positive $a_\pm$ and real $\delta_\pm$, and introduce the Stokes parameters \cite{15}
\begin{eqnarray}
\label{4.17}I&=&a_+^2+a_-^2,\qquad Q=a_+^2-a_-^2,\nonumber\\
U&=&2a_+a_-\cos\delta,\qquad V=2a_+a_-\sin\delta,
\end{eqnarray}
where $\delta$ is the phase difference $\delta_+-\delta_-$. Then Eq. (4.13) becomes
\begin{equation}
\label{4.18}\overline{U}_2=\;\frac{m}{\sqrt{(l+2)(2l+1)}}V\omega a,
\end{equation}
 The power $P=\overline{\mathcal{D}_2}$ is proportional to $I$ and the rate of rotation $|\overline{\Omega}_2|$ is a linear combination of $I$ and $Q$. For given amplitudes $a_\pm$, i.e. for given power and rate of rotation, the translational swimming speed $|\overline{U}_2|$ is maximal for phase difference $\delta=\pm\pi/2$. The translational swimming velocity can be positive or negative, depending on the sign of $m$ and $\sin\delta$. For fixed power and ratio $a_+/a_-$ the rate of rotation is maximal for $m=\pm l$. For fixed power and phase difference $\delta=\pm\pi/2$ the swimmer can vary its mean rate of rotation in such a way that the mean translational swimming velocity in the up or down direction is maximal.

\section{\label{V}Spinning Volvox}

Volvox can swim in a large variety of ways depending on the superposition of modes. The modes we considered in the preceding section are rather special, but nonetheless we can draw some tentative conclusions. In experimental observation Volvox is seen to be swimming with an axial stroke \cite{9}-\cite{11}, as first studied by Lighthill \cite{2} and Blake \cite{3}. In our notation the modes are of type $\vc{v}_{l0}$ and $\vc{u}_{l0}$, as given by Eq. (3.1). The flow pattern seen experimentally by velocimetry corresponds to a superposition of modes with approximately $l=3,4,5$. The corresponding mean translational swimming velocity and mean rate of dissipation can be evaluated from the explicit expressions for the $\du{A},\du{B}$-matrices in $(VU,m=0)$-representation given in earlier work \cite{8}. In the axial swimming mode the mean rotational velocity vanishes. In the experiment the Volvox is held fixed by micropipette and no azimuthal component of the flow pattern is observed. The flow pattern is measured at distance $r=1.3a$, so that short range effects can go unnoticed. The radius of the Volvox is approximately $a\approx 200\mu\mathrm{m}$ and the beat frequency $f=\omega/2\pi$ is about $32\;Hz$.

We can assume in particular that the superposition of modes contains a $TW$-mode contribution $(\vc{t}_{lm},\vc{w}_{lm})$, besides axial modes $(l'0)$. The corresponding mean swimming velocities are then
\begin{equation}
\label{5.1}\overline{U}_2=\overline{U}_{2A}+\overline{U}_{2R},\qquad\overline{\Omega}_2=\overline{\Omega}_{2R},
\end{equation}
where $\overline{U}_{2A}$ is the contribution of the axial modes to the mean swimming velocity, and $\overline{U}_{2R}$ and $\overline{\Omega}_{2R}$ are given by Eqs. (4.13) and (4.14). Since these correspond to $m\neq 0$ there is no interference with the axial modes. The  additional power required can be evaluated from Eq. (4.15). It seems reasonable to assume the same beat frequency $\omega$. If the rotational modes correspond to a large value of $l$ then the flow pattern is of quite short range, falling off as $(a/r)^l$ for the $\vc{t}_{lm}$-mode, as can be seen from Eqs. (3.2) and (4.1). The choice of pitch $m/l$, amplitudes $a_\pm$, and phase $\delta$ allows a wide range of values for the resulting translational and rotational swimming velocities.

It follows from Eqs. (4.14) and (4.15) that if either of the modes $\vc{t}_{lm},\;\vc{w}_{lm}$ is excited with the same power, then for the $\vc{t}_{lm}$ mode the rate of rotation is larger by a factor $(l^2+l-1)(l+2)/(2l+1)$ than for the mode $\vc{w}_{lm}$. If just one of the modes is excited, then according to Eq. (4.13) the translational swimming velocity $\overline{U}_{2R}$ vanishes. Volvox can swim with one rotational mode by use of axial modes $(l'0)$. Conceivably the number $l$ characterizing the spinning mode is much larger than the number $l'$ characteristic of the axial modes causing translational swimming.

A possible scenario is that Volvox preferably swims rotationally by mode $\vc{t}_{lm}$, since this requires least effort, and translationally by excitation of axial modes $(l'0)$. The beat frequency $\omega$ for both types of modes can be assumed to be the same. Possibly the multipole order $l$ of the rotational mode is related to the cell structure on the surface. If this is the case, then it is of the order of the number of cells on a circumference.

Ghose and Adhikari \cite{16} have made a detailed analysis of the experimental data for flows generated by Volvox which is held fixed. They find good agreement with the data for a set of axial modes of order $l=1,2,3$. They suggest that rotational swimming corresponds to a septlet mode, with flow proportional to $\hat{\vc{C}}_{30}(\theta)(a/r)^4$ in our notation. Such a mode describes a swirling azimuthal flow, but it does not correspond to net rotation, since
\begin{equation}
\label{5.2}\int^\pi_0\hat{\vc{C}}_{10}(\theta)\cdot\hat{\vc{C}}_{30}(\theta)\sin\theta\;d\theta=0.
\end{equation}
As we have discussed elsewhere \cite{17}, an analysis of swimming requires application of the no-slip boundary condition on a periodically deforming surface. It is not sufficient to postulate the time-dependent or time-averaged flow pattern. Since swimming in lowest order of the amplitude is a bilinear effect, one must also examine the interference of modes.

For example, for the rotational first order flow of Sec. III with moments $(0,0,\nu_{21},0)$ and complex flow velocity $\vc{v}^c_1=\nu_{21}\vc{w}_{21}(\vc{r})\exp(-i\omega t)$ there is no net rotation because of selection rules like Eq. (5.2),
\begin{equation}
\label{5.3}\int\hat{\vc{C}}^*_{1m}\cdot\hat{\vc{C}}_{21}\;d\Omega=0,\qquad (m=-1,0,1),
\end{equation}
but the body does acquire a rotational swimming velocity via the no-slip boundary condition on account of the running wave nature of the first order flow. In this situation the mean second order flow velocity $\overline{\vc{v}}_2$ has a contribution $-\overline{\vc{\Omega}}_2\times\vc{r}$ extending to infinity. It follows from Eq. (3.10) that the mean rotational swimming velocity is $\overline{\vc{\Omega}}_2=-\frac{9}{10}\;|\nu_{21}|^2\omega\;\vc{e}_z$.

\section{\label{VI}Discussion}

In the above we have suggested a possible scenario for the rotational swimming of Volvox. The tangential modes responsible for rotation may be of high multipole order, corresponding to a short range flow pattern. We showed that two tangential helical waves are of particular interest and suggested that the mode with nonvanishing pressure disturbance may be preferred by Volvox, because it requires less power for the same rate of rotation.

We presume that the tangential, rather than radial, helical waves are most relevant for swimming Volvox. Ehlers and Koiller \cite{18} have performed an approximate calculation of swimming velocities and power for transverse helical waves with displacement vector perpendicular to a spheroidal surface. They employ a tangent plane approximation and Taylor's results for a waving planar sheet \cite{19}, and suggest that their calculation is relevant to the swimming of $Synechococcus$. For a sphere the swimming by means of a radial helical wave could be studied more accurately by the method employed here. It would be of interest to compare the swimming efficiency of the radial helical wave with that of the tangential helical waves.

Leshansky et al. \cite{20} have studied the swimming of a spheroid by tanktreading motion of its surface. An extension of the present study to spheroidal geometry is challenging and  would be of interest.

\newpage

\section*{Figure captions}

\subsection*{Fig. 1}
Plot of the body surface at times $t=0,\;T/8,\;T/4$ for the screwing motion described at the end of Sec. III.

\newpage


\begin{thebibliography}{99}

\bibitem{1}N. Ueki, S. Matsunaga, I. Inouye, and A. Hallman, How 5000 independent rowers coordinate their
strokes in order to row into the sunlight: Phototaxis in the multicellular green alga Volvox, BMC Biology \vol{8}, 103 (2010).

\bibitem{2}
M. J. Lighthill, On the squirming motion of nearly spherical deformable bodies through liquids at very small Reynolds numbers, Comm. Pure Appl. Math. \vol{5}, 109 (1952).

\bibitem{3}
J. R. Blake, A spherical envelope approach to ciliary propulsion, J. Fluid Mech. \vol{49}, 209 (1971).

\bibitem{4}
A. Shapere and F. Wilczek, Geometry of self-propulsion at low Reynolds number, J. Fluid Mech. \vol{198}, 557 (1989).

\bibitem{5}
A. Shapere and F. Wilczek, Efficiencies of self-propulsion at low Reynolds number, J. Fluid Mech. \vol{198}, 587 (1989).

\bibitem{6}
B. U. Felderhof and R. B. Jones, Inertial effects in small-amplitude swimming of a finite body, Physica A \vol{202}, 94 (1994).

\bibitem{7}
B. U. Felderhof and R. B. Jones, Small-amplitude swimming of a sphere, Physica A \vol{202}, 119 (1994).

\bibitem{8}
B. U. Felderhof and R. B. Jones, Optimal translational swimming of a sphere at low Reynolds number, Phys. Rev. E \vol{90}, 023008 (2014).

\bibitem{9}
K. Drescher, R. E. Goldstein, N. Michel, M. Polin, and I. Tuval, Direct measurement of the flow field around swimming microorganisms, Phys. Rev. Lett. \vol{105}, 168101 (2010).

\bibitem{10}
D. R. Brumley, P. Polin, T. J. Pedley, and R. E. Goldstein, Hydrodynamic synchronization and metachronal waves on the surface of the colonial alga $Volvox\;carteri$, Phys. Rev. Lett. \vol{109}, 268102 (2012).

\bibitem{11}
D. R. Brumley, P. Polin, T. J. Pedley, and R. E. Goldstein, Metachronal waves in the flagellar beating of $Volvox$ and their hydrodynamic origin, Phys. Rev. Lett. \vol{109}, 268102 (2012).

\bibitem{12}
B. Cichocki, B. U. Felderhof, and R. Schmitz, Hydrodynamic interactions between two spherical particles, PhysicoChem. Hyd. \vol{10}, 383 (1988).

\bibitem{13}
A. R. Edmonds, {\it Angular Momentum in Quantum Mechanics} (Princeton University Press, Princeton, NJ, 1974).

\bibitem{14}
B. U. Felderhof and R. Schmitz, Creeping flow about a sphere, Physica A \vol{92}, 423 (1978).

\bibitem{15}
C. F. Bohren and D. R. Huffman, {\it Absorption and Scattering of Light by Small Particles}, (Wiley, New York, 1983).

\bibitem{16}
S. Ghose and R. Adhikari, Irreducible representation of oscillatory and swirling flows in active soft matter, Phys. Rev. Lett. \vol{112}, 118102 (2014).

\bibitem{17}
B. U. Felderhof, Stokesian spherical swimmers and active particles, Phys. Rev. E \vol{91}, 043018 (2015).

\bibitem{18}
K. M. Ehlers and J. Koiller, Micro-swimming without flagella: propulsion by internal structures, Regular and Chaotic Dynamics \vol{16}, 623 (2011).

\bibitem{19}
G. I. Taylor, Analysis of the swimming of microscopic organisms, Proc. R. Soc. London A \vol{209}, 447 (1951).

\bibitem{20}
A. M. Leshansky, O. Kenneth, O. Gat, and J. E. Avron, A frictionless microswimmer, New J. Phys. \vol{7}, 145 (2007).
\end{thebibliography}
\end{document}